\documentclass[twocolumn,usenatbib]{mn2e}
\usepackage{aas_macros}
\usepackage{subfigure}
\usepackage{graphicx}
\usepackage{psfig}

\title[The radio structure of UHBLs]{The radio structure of ultra-high-energy synchrotron peak BL Lacs}
\author[Z. Z. Wu, D. Jiang and M. Gu]
{Zhongzu Wu$^{1}$\thanks{E-mail: zzwu08@gmail.com}, D. R.
Jiang$^{2}$, Minfeng Gu$^{2}$
\\
\\
$^{1}$ College of Science, Guizhou University, Guiyang 550025, China.
\\
$^{2}$ Key Laboratory for Research in Galaxies and Cosmology,
Shanghai Astronomical Observatory, Chinese Academy of Sciences,\\ 80
Nandan Road, Shanghai 200030, China
\\
$^{}$
}

\begin{document}

\pagerange{\pageref{firstpage}--\pageref{lastpage}} \pubyear{2012}

\maketitle


\begin{abstract}
We present the results of EVN and MERLIN 5 GHz observations of nine
ultra-high-energy synchrotron peak BL Lacs (UHBLs) selected as all
BL Lacs with \textbf{log ($\nu_{\rm peak}/ \rm Hz)>20$}  from Nieppola et al.. The radio
structure was investigated for these sources, in combination with
the available VLBA archive data. We found that the core-jet
structure is detected in five sources, while four sources only have
a compact core on pc scale. The core of all sources shows high
brightness temperature (with mean and median values
\textbf{log ($T_{\rm b} / {\rm K}) \sim11$} , which implies that the beaming effect likely present in
all sources. When the multi-epoch VLBI data are available, we found no
significant variations either for core or total flux density in two
sources (2E 0414+0057 and EXO 0706.1+5913), and no evident proper
motion in 2E 0414+0057, while the superluminal motion is likely detected in EXO 0706.1+5913. Our sources are found to be less compact
than the typical HBLs in Giroletti et al, by comparing the ratio of
the VLBI total flux to the core flux at arcsec scale. Combining all
our results, we propose that the beaming effect might be present in
the jets of UHBLs, however, it is likely weaker than that of typical
HBLs. Moreover, we found that UHBLs could be less Doppler beamed
versions of HBLs with similar jet power, by comparing the
distribution of redshift, and radio luminosities. The results are in
good consistence with the expectations from our previous work.
\end{abstract}

\section{Introduction}
BL Lac objects are a type of radio loud active galactic nuclei
(AGNs) with no emission lines or less of them and characterized by
the nonthermal radiation over the entire electromagnetic spectrum.
They can be classified as different subclasses based on their
spectral energy distribution (SED), namely, low frequency peaked BL
Lac objects (LBL), intermediate objects (IBL) and high frequency
peaked BL Lac objects (HBL) \citep{padovani95}. Generally, LBLs are
intrinsically more luminous than HBLs with the 5 GHz radio
luminosity, $\gamma$-ray luminosity and the luminosity at peak
frequency $\nu_{\rm peak}$ of the synchrotron component, which is
called blazar sequence \citep[see][for details]{fossati98,ghisel98}.

The power - $\nu_{\rm peak}$ anti-correlation suggested by blazar
sequence has been tested by larger blazar samples in recent years;
however, the results are controversial \citep{padovani07}.
\cite{meyer11} shows that blazar sequence is formed from two
populations in the synchrotron $\nu_{\rm peak}$ - $L_{\rm peak}$
plane, each forming an upper edge to an envelope of progressively
misaligned blazars and most intermediate synchrotron peak (ISP)
sources are not intermediate in intrinsic jet power between LSP and
high synchrotron-peaking (HSP) sources, but are more misaligned
versions of HSP sources with similar jet powers. \cite{rani2011}
found that the changes in the jet Doppler factor is the most
important one among the model parameters responsible for changes
in the observed SEDs of blazars. \cite{wu06} found a significant
anti-correlation between the intrinsic $\nu_{\rm peak}$ and the
total luminosity at 408 MHz, which is less affected by the beaming
effect thus represents the intrinsic radio emission. As a whole,
LBLs are found to be more powerful than HBLs. The authors claimed
that HBLs have smaller Doppler factor and larger viewing angle than
LBLs. Intriguingly, after eliminating the beaming effect in both
parameters, \cite{nieppola2008} and \cite{wu09} found a positive
correlation between the synchrotron peak frequency and luminosity,
which is opposite to the blazar sequence.

In order to understand the differences between the subclasses of BL
Lacs, the high resolution radio observations have been performed to
explore the jet properties for HBLs and LBLs, and the comparison
between these two populations were made. The VLBI observations
showed that most LBLs display superluminal motions
\citep{chen99,Gabuzda00,jorstad01}, while the TeV blazars (most are
HBLs) display subluminal or mildly relativistic motions
\citep{gir04b,gir06,pin04}. \cite{rec03} observed 15 HBLs and 3 RGB
BL Lacs, and found that HBLs, like most LBLs, show parsec-scale
core-jet morphologies with complex kilo-parsec scale morphologies.
Moreover, the jets in HBLs are more well aligned, suggesting that
their jets are either intrinsically straighter or are seen further
off-axis than LBLs. \cite{gir04a} selected 30 low redshift BL Lac
objects and confirmed that HBLs show less distortion and therefore,
are expected to be oriented at larger angles than LBLs. All these
findings seem to be consistent with the results of \cite{wu06}, of
which HBLs have larger viewing angles than that of LBLs.

\cite{Ghise99} suggested that there is a class of BL Lacs with the
synchrotron peak at higher frequencies than that of conventional
HBLs, i.e. \textbf{log ($\nu_{\rm peak}/ \rm Hz)>19$}, and these sources can be
called ultra-high-energy synchrotron peak BL Lacs (UHBLs)
\citep{Giommi01}. As these sources are at the extreme end of
$\nu_{\rm peak}$ distribution, the investigation of radio compact
structure, and jet property are thus crucial in studying BL Lacs
populations. As \cite{wu06} suggested, UHBLs are expected to have
smaller Doppler factor, larger viewing angle, and lower radio
luminosity. However, as far as we know, the VLBI observations are
only presented for a few UHBLs, and the radio compact structures of
UHBLs are largely unknown. In this paper, we investigate the radio
structure of UHBLs based on our EVN and MERLIN observations and VLBI
archive data.

The layout of this paper is as follows. In Section 2, the
observations and data reduction are presented. The results are
described in Section 3. The discussions are shown in Section 4,
while the conclusions are drawn in Section 5. Throughout the paper,
we define the spectral index $\alpha$ as $\rm
f_{\nu}\propto\nu^{-\alpha}$, where $f_{\nu}$ is the flux density at
frequency $\nu$, and a cosmology with $\rm H_{0}=70 \rm {~km ~s^
{-1}~Mpc^{-1}}$,
$\rm \Omega_{M}=0.3$, $\rm \Omega_{\Lambda} = 0.7$ is adopted. 

\begin{table*}
\caption{The VLBI observation log.} \label{table_log}
\begin{tabular}{lllllll}
\hline\hline

Object & $z$ &Epoch &Freq &Array \\
       & &(years)   & (GHz)     & \\\hline
          2E 0414+0057&0.287& 1996.68   &    4.99  &VLBA        \\
                       && 1997.38   &    4.99  &VLBA        \\
                       && 1999.70   &    4.99  &VLBA        \\
                       && 2004.73   &    4.99  &VLBA        \\
                       && 2004.73   &    8.42  &VLBA        \\
                       && 2009.16   &    4.99  &EVN+MERLIN  \\
                       && 2010.99   &    8.65  &VLBA        \\
                       && 2011.02   &    8.65  &VLBA        \\
 EXO 0706.1+5913       &0.125& 2002.13   &    5.00  &VLBA        \\
                       && 2004.73   &    4.99  &VLBA        \\
                       && 2004.73   &    8.42  &VLBA        \\
                       && 2009.16   &    4.99  & EVN+MERLIN \\
                       && 2010.93   &    8.65  &VLBA        \\
 1ES 0927+500          &0.14& 2009.16   &    4.99  & EVN+MERLIN \\
                       && 2010.15   &    8.42  &VLBA        \\
                       && 2010.15   &   15.37  &VLBA        \\
   RXS J1012.7+4229    &0.364& 2009.16   &    4.99  & EVN+MERLIN \\
  RGB 1319+140         &0.573& 2009.16   &    4.99  & EVN+MERLIN \\
  RXS J1341+3959       &0.163& 2009.16   &    4.99  &EVN+MERLIN  \\
                       && 2009.79   &    8.42  &VLBA        \\
                       && 2009.79   &   15.37  &VLBA        \\
  RXS J1410+6100       &0.384& 2009.16   &    4.99  &EVN+MERLIN  \\
 RXS J1458.4+4832      &0.541& 2009.16   &    4.99  &EVN+MERLIN  \\
RXS J2304.6+3705       &0.57& 2009.16   &    4.99  &EVN+MERLIN  \\

\hline

\end{tabular}
\end{table*}
\begin{table*}
 \caption{The EVN+MERLIN image parameters.}
\label{table_img}
\begin{tabular}{lllllll}
\hline\hline

Object &  \textbf{ $z$  } &HPBW & Noise ($3\sigma$) & Peak & Array\\
       &        &(mas $\times$ mas,$^\circ$)&(mJy beam$^{-1}$)&(mJy beam$^{-1}$)
&

\\\hline

1ES 0927+500& 0.14       &  $2.63 \times 1.61,\, 22.2$ &  0.18  &12.4      &  1     \\
          &         &  $63.3 \times 41.2,\, -62.8$   &  1.06   &23.7      &  2          \\
          &          &  $2.75 \times 1.69,\, 22.8$  &  0.17  &12.6        &  1+2         \\

RXS J1012.7+4229&  0.364     &  $4.7 \times 2.49,\, -3.4$   &  0.15   &19.6       &  1          \\
               &        &  $62.4 \times 42.3,\, 43.1$   &  1.04   &32.6       &  2          \\
               &       &  $4.9 \times 3.62,\, -9.27$   &  0.31   &19.9      &  1+2          \\

RGB 1319+140& 0.573   &  $6.23 \times 2.1, \,15.2$   &  0.263   &26.2       &  1      \\
                &      &  $70.4 \times 49.7, \,23.5$   &  1.4   &41.7       &  2      \\
             &     &  $6.35 \times 2.23, \,15.6$   &  0.302   &32.2       &  1+2      \\

RXS J1341+3959& 0.163  &  $5.82\times 2.23, \,30.7$   &  0.25   &6.9       &  1     \\
                &          &  $54.4 \times 46.6, \,41$   &  0.979  &22.0       &  2         \\

RXS J1410+6100& 0.384    &  $2.79 \times 2.07, \,2.23$   &  0.17  &7.15      &  1     \\
                    &         &  $53.9 \times 48.1, \,-79.8$  &  0.81  &17.6       &  2          \\

  RXS J1458.4+4832        &   0.541      &  $4.19 \times 2.28, \,30.2$   &  0.168  &7.7      &  1          \\
                        &        &  $52.6 \times 46.5, \,29.9$   &  0.844  &18.4      &  2          \\
  RXS J2304.6+3705        &   0.57     &  $4.89 \times 2.19, \,0.999$  &  0.092  &9.72      &  1         \\
                          &          &  $63.6 \times 40.6, \,47.5$  &  1.12  &22.2      &  2         \\

\hline\hline
\end{tabular}
\vskip 0.1 true cm \noindent  1: EVN, 2: MERLIN.
\end{table*}

\section{Observation and data reduction}
Nieppola et al. (2006) have
constructed the SEDs for a large, heterogeneous sample of BL Lacs taken from the Veron-Cetty \& Veron BL Lac catalogue and
visible from the Metsh$\ddot{a}$hovi radio observatory. This is the first time the SEDs of BL Lacs have been studied
with a sample of over 300 objects.
 In the sample, 22 BL
Lacs with $\nu_{\rm peak}>10^{19}$ Hz were classified as UHBLs
candidates. From these sources, we selected all nine sources with
\textbf{log ($\nu_{\rm peak}/ \rm Hz)>20$}, which represent the extreme population of
UHBLs. In order to explore their radio structure, the VLBI
simultaneous observations with EVN and MERLIN, were carried out at 5
GHz for these sources in February 2009 with a total observing time
of 24 hours. The phase reference technique was used for all sources
except for 2E 0414+0057 and EXO 0706.1+5913. The phase calibrators
were found in the VLBA calibrators, whose distances from the targets
are less than 3 degrees. In order to get higher angular resolution,
we request also two Chinese telescopes, but unfortunately, the
Urumqi station was unavailable during the observation. The recording
rate of 1 Gb/s was adopted, and the scan time around 1.5 hours was
taken for each source.

Besides our observations, we also collected the available VLBA
archive data. All these observations are listed in Table
\ref{table_log}. The data reduction were performed using the NRAO
Astronomical Image Processing System (AIPS). The imaging and model
fitting were carried out with DIFMAP package (Shepherd, Pearson \&
Taylor 1994). The image parameters for our EVN and MERLIN
observations are listed in Table \ref{table_img}: Col. (1) source
name, Col. (2) redshift, Cols. (3) - (4) the observational date and
the half-power beamwidth (HPBW) of the weighted beam, Col. (5) the
noise of the image, Col. (6) the peak flux of the image, Col. (7)
VLBI array.



\begin{figure*}

  \begin{center}
    \includegraphics[width=0.6\textwidth,angle=360]{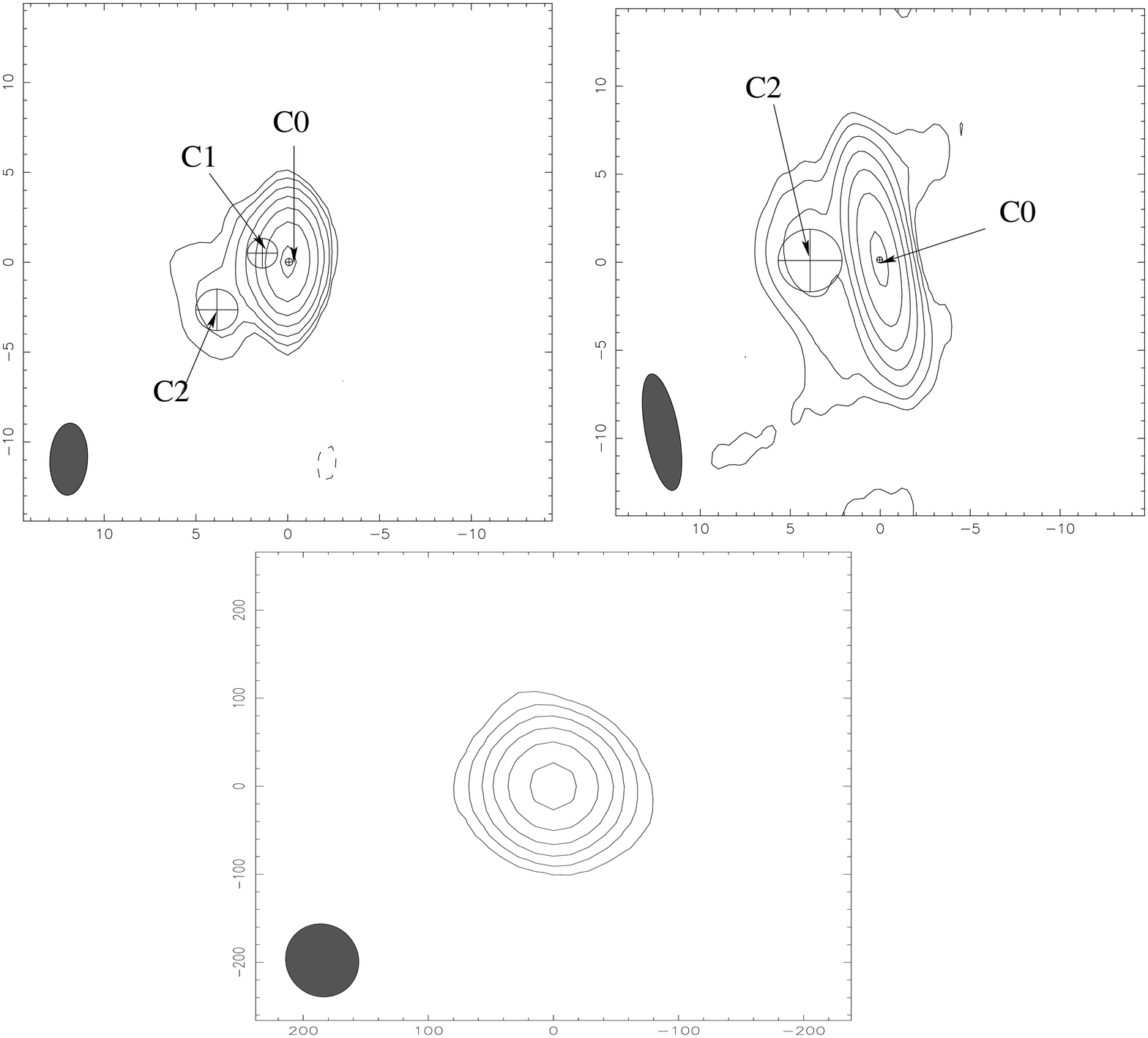}

  \end{center}
\caption{The radio structure of 2E 0414+0057: upper left - VLBA at 5 GHz, epoch 1997.38;
upper right - our EVN observation; bottom: Our MERLIN observation. The core, and two jet components are
labeled with C0, C1 and C2, respectively. The contour levels are
(-1, 1, 2, 4, 8, 16, 32, 64) multiples the $3\sigma$ level. The
horizontal axis is the relative R. A., and the vertical axis is the
relative Dec. to the source position in mas. \label{J0414com}}
\end{figure*}

\section{Results}

The radio components from the circular gaussian model fitting for EVN images are
listed in Table \ref{table_model}, in which Col. (1) is the source
name, Col. (2) the observation epoch, Col. (3) the observation
frequency, Col. (4) the label of radio components, Col. (5) the flux
density and uncertainties of corresponding components (in mJy), Col.
(6) the distance of component to core (in mas), Col. (7) the
position angle of component, Col. (8) the radius of components, and
Col. (9) the brightness temperature (see Section 4.1). All the
uncertainties of the component parameters were derived using the
formula given by
Andrei Lobanov  \footnote{http://www.radionet-eu.org/rda/archive/eris-11$_{-}$lobanov.pdf}.

\subsection{Source position}

With the phase-reference technique, we are able to obtain the
accurate positions from the VLBI data, which are listed in Table
\ref{table_position} for seven sources. All the celestial positions
we used for correlation are from NASA/IPAC Extragalactic Database
(NED)\footnote{http://ned.ipac.caltech.edu/}. Our results show that
the accurate position for RXS J1341+3959 and RXS J1410+6100 are
significantly different from the positions provided by NED, with
position difference larger than $10^{''}$. The accurate positions
for these two sources can only be derived from our MERLIN data (see
Table \ref{table_position}), which however, are much closer to that
from the Faint Images of the Radio Sky at Twenty Centimeters (FIRST)
1.4-GHz radio catalogue \citep{bec95}, with a position difference
less than $0^{''}.3$ and $0^{''}.5$, respectively. The accurate
positions of other sources are obtained from the EVN data, which
might be more accurate than those from the FIRST images.

\subsection{Individual sources}

\subsubsection{2E 0414+0057}
The radio structure has been studied by \cite{rec03} and
\cite{kharb08}. The 5 GHz VLA image of this source shows a jet like
feature at P.A. $\sim40^{\circ}$, while the VLBI images show a jet
in P.A. $\sim75^{\circ}$ \citep{kharb08}. The VLBA map resolves a
jet that initially extends to the east-northeast (P.A.
$\sim68^{\circ}$) of the core and also weak, extended ($\sim$ 3 - 4
$\sigma$) emission to the southeast of the jet, which suggests
either that the jet is collimated and bends to the south $\sim10$ pc
from the core, or that the projected jet opening angle is wide
($\sim60^{\circ}$). The inner portion of the jet is well aligned
($\Delta$P.A. $= 5^{\circ}$) to the kpc-scale jet \citep{rec03}.

The radio structure of our EVN and MERLIN observation for this source is shown
in Fig. \ref{J0414com}, as well as one VLBA image as comparison.
There are two jet components in VLBA image, however, one of them is
absent in our EVN image, while the MERLIN image shows only a compact core. All the measured parameters for core and
jet components are listed in Table \ref{table_model}, from which it
can be seen that the P.A. of inner jet component ranges from
56$^{\circ}$ to 90$^{\circ}$, while it is 88$^{\circ}$ -
124$^{\circ}$ for the outer one. Therefore, our results also suggest
that the jet of this source has a wide opening angle. Based on the
multi-epoch VLBI data, we calculated the spectrum index of core with
average flux at 4.9 and 8.6 GHz, and found it is with
$\alpha=0.26$.

\subsubsection{EXO 0706.1+5913}
The 1.4 GHz VLA image shows that the radio core is located on the
north-west edge of a roughly spherical cocoon. It was suggested that
this object could be a wide-angle or narrow-angle tailed object,
viewed at a small angle. The 5 GHz VLBA image shows a weak jet to
the south-west \citep{gir04a}.

\begin{figure*}

  \begin{center}
    \includegraphics[height=.3\textheight]{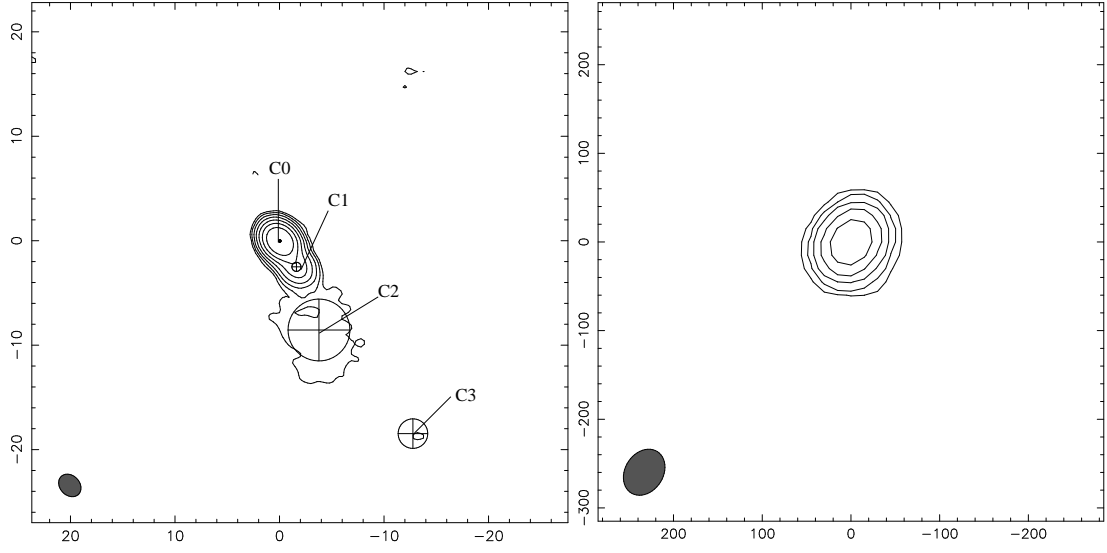}
  \end{center}
\caption{The radio structure of EXO 0706.1+5913 from our EVN and MERLIN
observation: left - our EVN observation; right our MERLIN observation. The core and three jet components are marked in the
image. The contour levels are (1, 2, 4, 8, 16, 32, 64) multiples the
$3\sigma$ level. The horizontal axis is the relative R. A., and the
vertical axis is the relative Dec. to the source position in mas.
\label{J0710com}}
\end{figure*}

We present our EVN and MERLIN image of this source in Fig. \ref{J0710com}. The
core and three jet components are detected from EVN observation, and their parameters are
listed in Table \ref{table_model} for all the VLBI data. We found
that the P.A. of jet components from all the VLBI data ranges from
-128$^{\circ}$ to -170$^{\circ}$, indicating that the projected jet
opening angle is likely around 42$^{\circ}$. A flat spectra is found
with $\alpha$=0.22 for the core, using the average flux at 4.9 and
8.6 GHz from multi-epoch VLBI data (see Table \ref{table_model}).

\subsubsection{1ES 0927+500}
\begin{figure*}

  \begin{center}
    \includegraphics[height=.5\textheight]{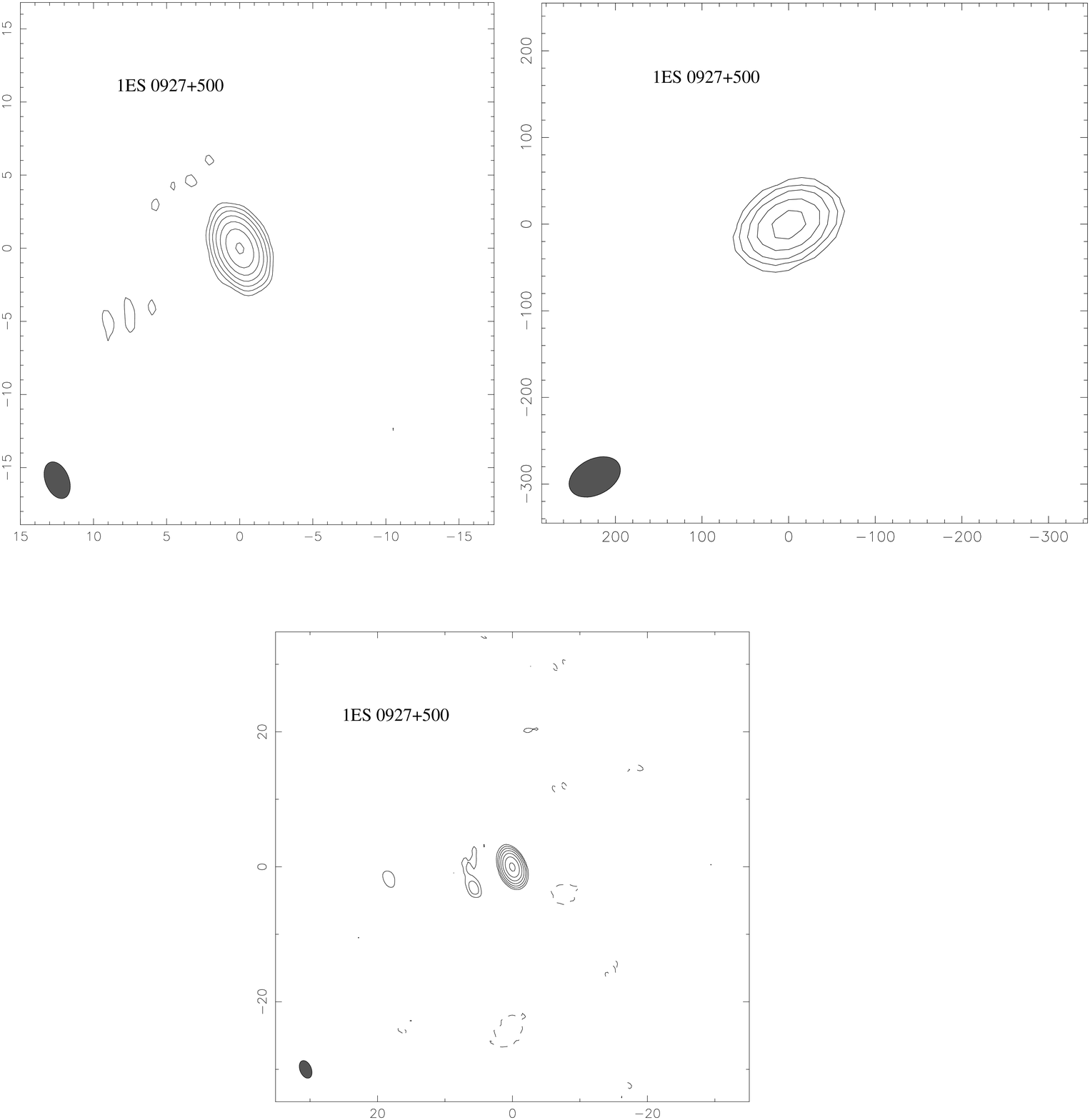}
  \end{center}
\caption{The EVN (upper left), MERLIN (upper right) and EVN+MERLIN
(bottom) images of 1ES 0927+500 at 5 GHz. The contour levels are
(-1, 1, 2, 4, 8, 16, 32, 64...) multiples the minimum contour level,
which is 3 times the rms noise given in Table \ref{table_img}. The
horizontal axis is the relative R. A., and the vertical axis is the
relative Dec. to the source position in mas. \label{J0930em}}
\end{figure*}

This source was observed with VLA config A at 1.4 GHz, with a peak flux density
19.9 mJy and no information is available on the pc-scale morphology
\citep{gir04a}. Our EVN, MERLIN and EVN+MERLIN images (Fig.
\ref{J0930em}) shows that this source has a compact core, and the
weak diffuse emission at south-east to the core is detected at EVN
and EVN+MERLIN images. The measured core parameters are shown in
Table \ref{table_model}.
\subsubsection{RXS J1012.7+4229}
\begin{figure*}
  \begin{center}
    \includegraphics[height=.5\textheight]{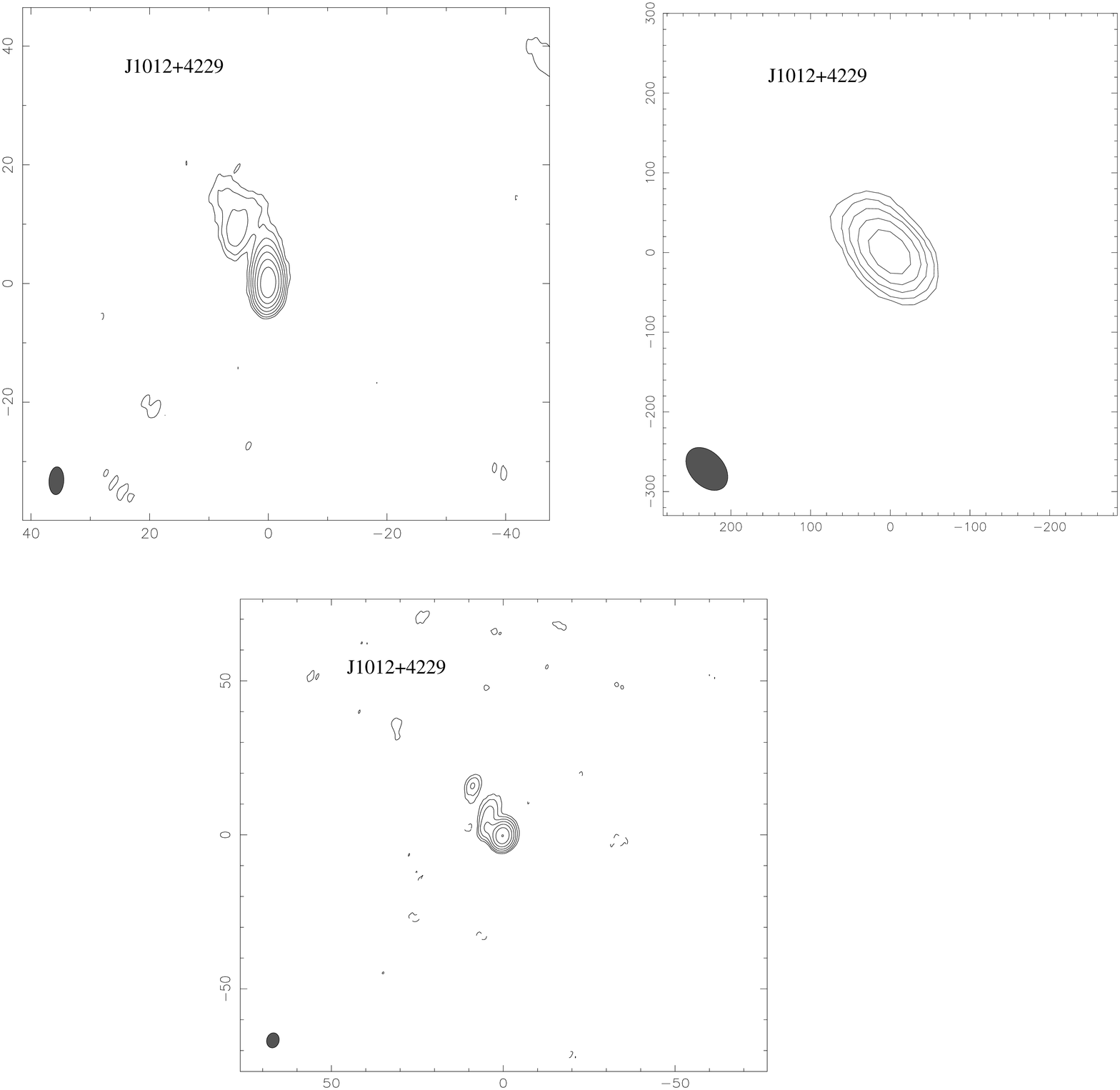}
  \end{center}
\caption{The EVN (upper left), MERLIN (upper right) and EVN+MERLIN
(bottom) images of RXS J1012.7+4229 at 5 GHz. The contour levels are
(-1, 1, 2, 4, 8, 16, 32, 64...) multiples the minimum contour level,
which is 3 times the rms noise given in Table \ref{table_img}. The
horizontal axis is the relative R. A., and the vertical axis is the
relative Dec. to the source position in mas. \label{J1012em}}
\end{figure*}

The total flux at 1.4 GHz for this source is around 79 mJy from both
FIRST and the NRAO VLA Sky Survey (NVSS) . Our EVN, MERLIN and
EVN+MERLIN images are shown in Fig. \ref{J1012em}. The EVN,
EVN+MERLIN images show that there is a jet bending to the northeast
direction. Consistently, there is a hint of jet structure towards
the northeast direction in the MERLIN image. Two jet components are
found with P.A. around 6$^{\circ}$ and 25$^{\circ}$, respectively,
which is shown in Table \ref{table_model}.
\subsubsection{RXS J1319+1405}
\begin{figure*}
  \begin{center}
    \includegraphics[height=.5\textheight]{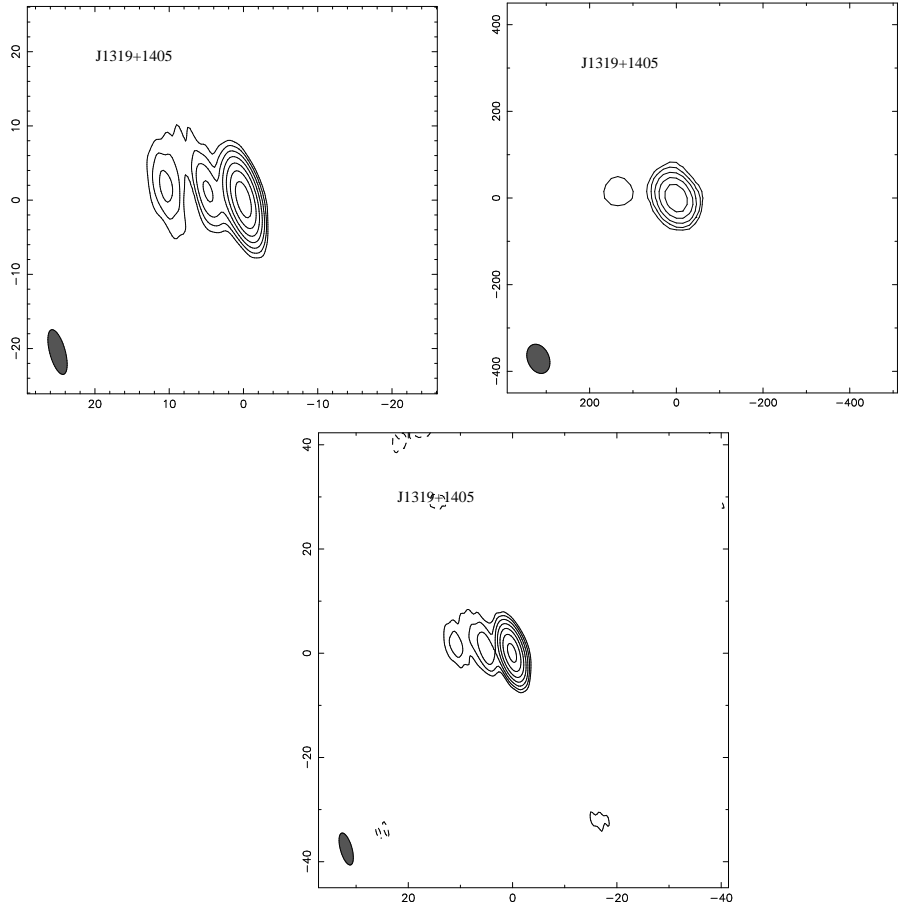}
  \end{center}
\caption{The EVN (upper left), MERLIN (upper right) and EVN+MERLIN
(bottom) images of RXS J1319+1405 at 5 GHz. The contour levels are
(-1, 1, 2, 4, 8, 16, 32, 64...) multiples the minimum contour level,
which is 3 times the rms noise given in Table \ref{table_img}. The
horizontal axis is the relative R. A., and the vertical axis is the
relative Dec. to the source position in mas. \label{J1319em}}
\end{figure*}

The total flux at 1.4 GHz for this source is around 65.8 mJy from
FIRST catalogue. Our EVN and MERLIN images are presented in Fig.
\ref{J1319em}. The EVN and EVN+MERLIN images of this source shows a
core-jet structure with jet towards east direction with P.A. around
80$^{\circ}$ (see Table \ref{table_model}). In the MERLIN image, the
sign of weak emission is found at much larger distance, but at
similar direction. The image parameters are shown in Table
\ref{table_img}.

\subsubsection{RXS J1341+3959}
\begin{figure*}
  \begin{center}
    \includegraphics[height=.3\textheight]{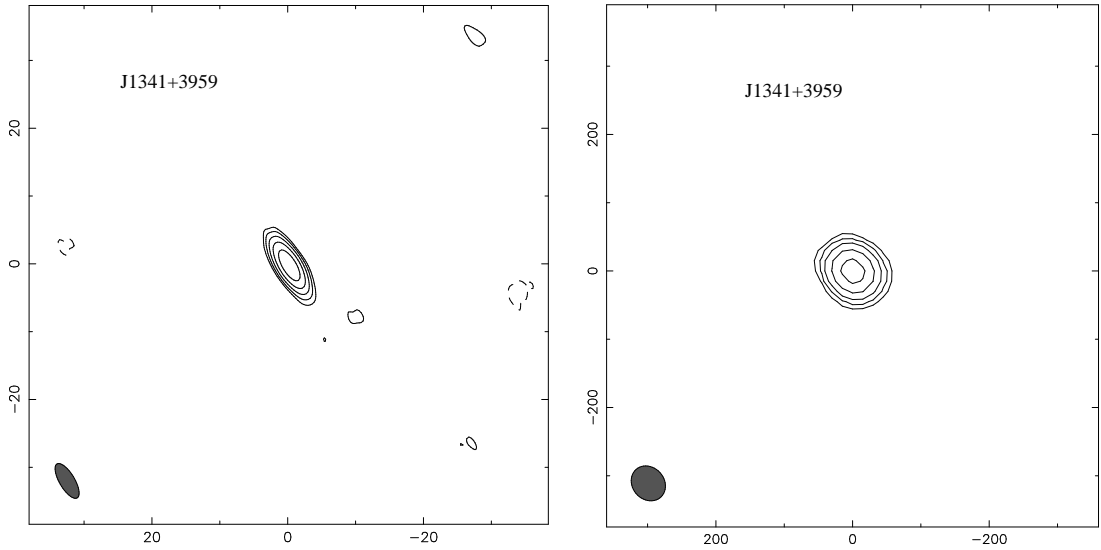}
  \end{center}
\caption{The EVN (left) and MERLIN (right) images of RXS J1341+3959
at 5 GHz. The contour levels are (-1, 1, 2, 4, 8, 16, 32, 64...)
multiples the minimum contour level, which is 3 times the rms noise
given in Table \ref{table_img}. The horizontal axis is the relative
R. A., and the vertical axis is the relative Dec. to the source
position in mas. \label{J1341em}}
\end{figure*}

The total flux at FIRST 1.4 GHz for this source is around 39.9 mJy.
Both the EVN and MERLIN images show a compact core, which is
presented in Fig. \ref{J1341em}. The image parameters are shown in
Table \ref{table_img}.

\subsubsection{RXS J1410+6100}
\begin{figure*}
  \begin{center}
    \includegraphics[height=.2\textheight]{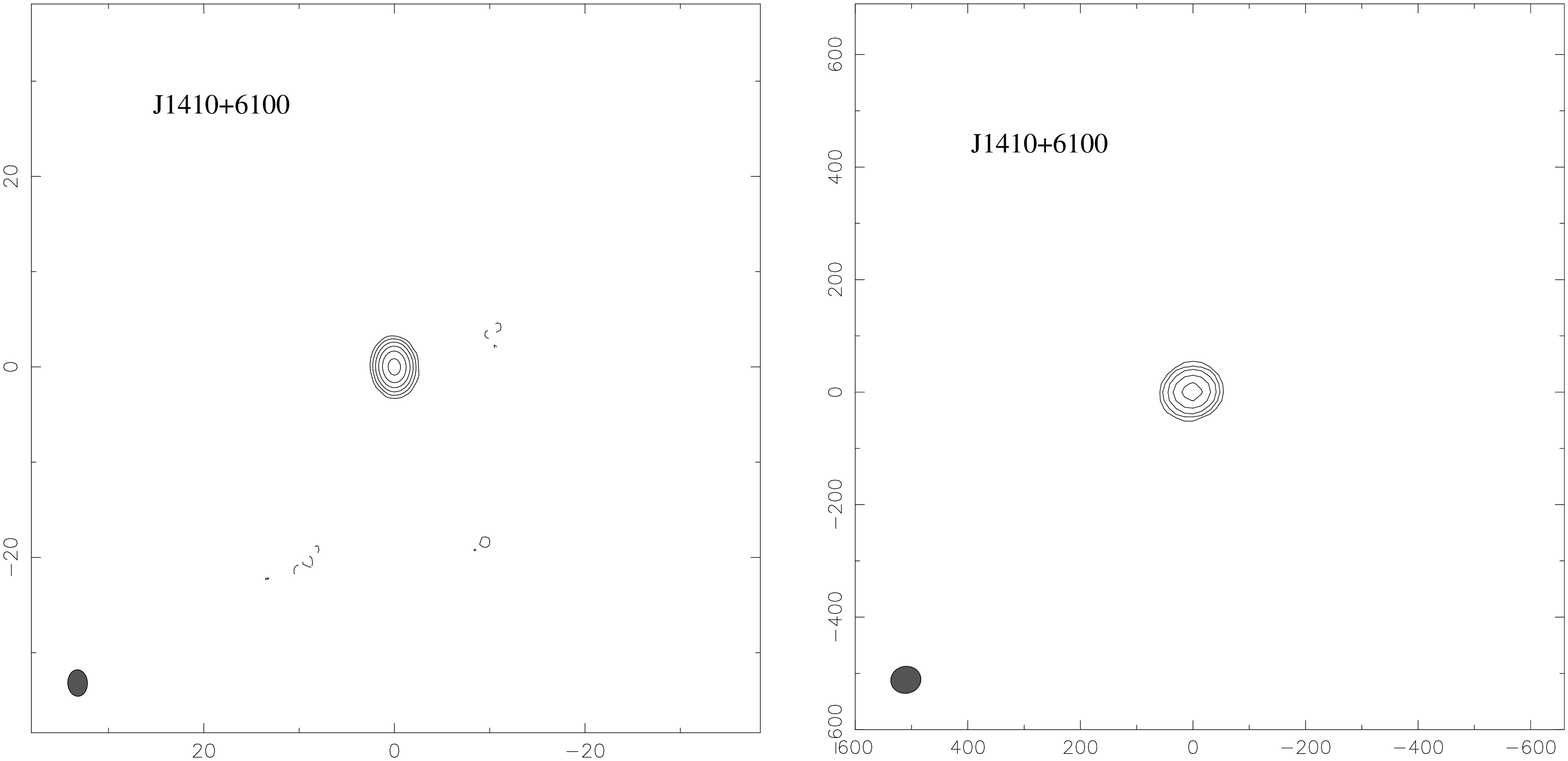}
  \end{center}
\caption{The EVN (left) and MERLIN (right) images of RXS J1410+6100
at 5 GHz. The contour levels are (-1, 1, 2, 4, 8, 16, 32, 64...)
multiples the minimum contour level, which is 3 times the rms noise
given in Table \ref{table_img}. The horizontal axis is the relative
R. A., and the vertical axis is the relative Dec. to the source
position in mas. \label{J1410em}}
\end{figure*}

The total flux at 1.4 GHz for this source is around 5.4 mJy from
VLA FIRST. Similar to RXS J1341+3959, a compact core is observed
from both the EVN and MERLIN images (Fig. \ref{J1410em}), with the
image parameters shown in Table \ref{table_img}. We found that the
EVN 5 GHz core flux density is about 7.2 mJy, which however is
higher than the FIRST VLA 1.4 GHz total flux. This indicates that
either the radio spectrum is invert, or there is variations in the
flux at these two radio bands.
\subsubsection{RXS J1458.4+4832}
\begin{figure*}
  \begin{center}
    \includegraphics[height=.2\textheight]{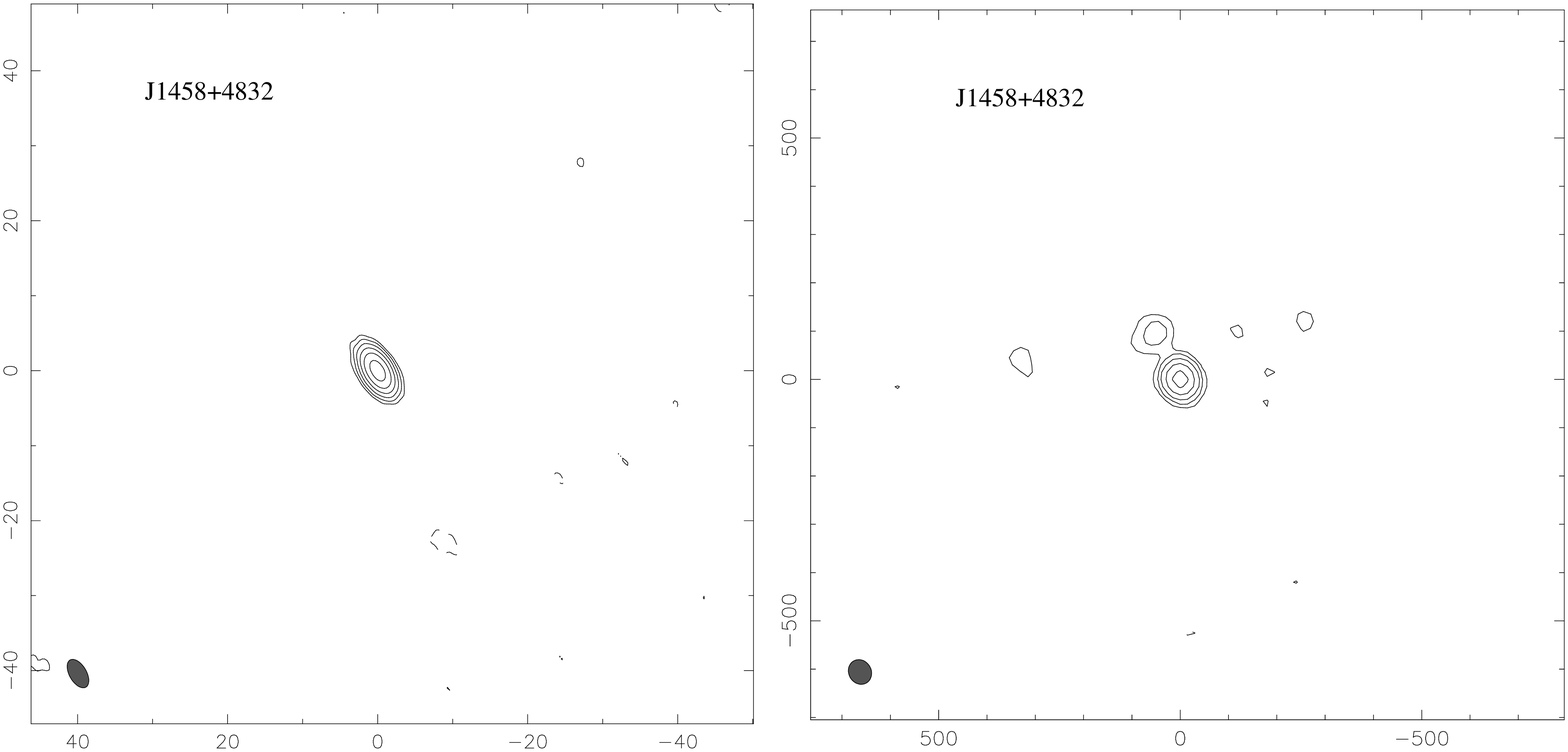}
  \end{center}
\caption{The EVN (left) and MERLIN (right) images of RXS
J1458.4+4832 at 5 GHz. The contour levels are (-1, 1, 2, 4, 8, 16,
32, 64...) multiples the minimum contour level, which is 3 times the
rms noise given in Table \ref{table_img}. The horizontal axis is the
relative R. A., and the vertical axis is the relative Dec. to the
source position in mas. \label{J1458em}}
\end{figure*}

The total flux at 1.4 GHz for this source is around 2.5 mJy from VLA
NVSS. We show our radio images in Fig. \ref{J1458em}. The EVN image
shows a compact core, while MERLIN image shows weak diffuse emission
at a large scale towards northeast direction. Similar to RXS
J1410+6100, the EVN 5 GHz total flux $\sim$7.6 mJy and the MERLIN
peak flux $\sim$18.4 mJy/beam are apparently larger than the total
NVSS flux, which implies either a invert spectrum or flux variation
in RXS J1458.4+4832.
\subsubsection{RXS J2304.6+3705}
\begin{figure*}
  \begin{center}
    \includegraphics[height=.2\textheight]{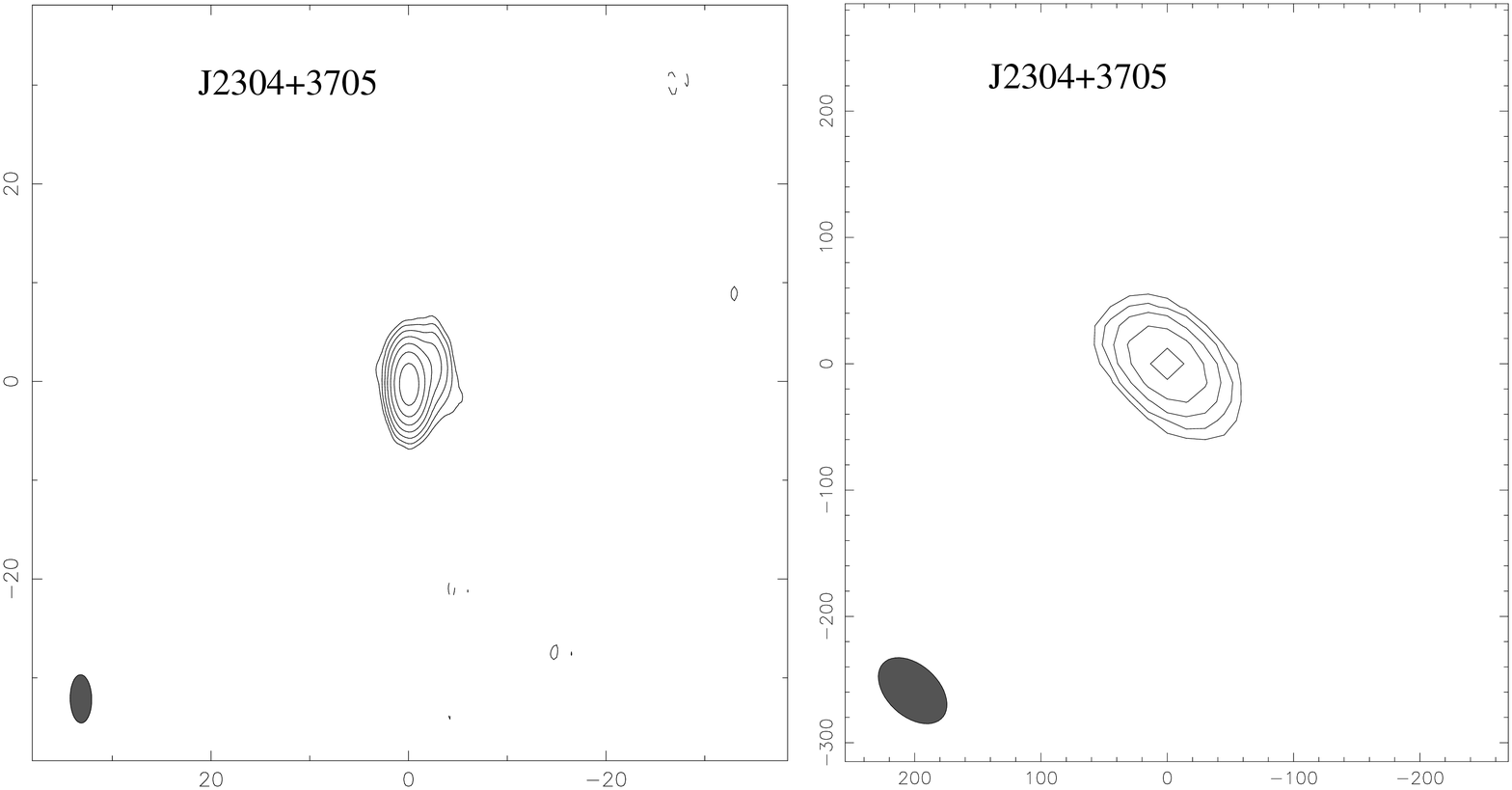}
  \end{center}
\caption{The EVN (left) and MERLIN (right) images of RXS
J2304.6+3705 at 5 GHz. The contour levels are (-1, 1, 2, 4, 8, 16,
32, 64...) multiples the minimum contour level, which is 3 times the
rms noise given in Table \ref{table_img}. The horizontal axis is the
relative R. A., and the vertical axis is the relative Dec. to the
source position in mas. \label{J2304em}}
\end{figure*}

The total flux at 1.4 GHz for this source is around 22.5 mJy from
VLA NVSS data. Our EVN image shows a possible jet structure towards
northwest direction with P.A. $\sim -60^{\circ}$ (see Fig.
\ref{J2304em} and Table \ref{table_model}). It has a total radio
flux around 11 mJy from EVN image. The MERLIN image shows only a
compact core, with peak flux about 22 mJy/beam.

\section{ Discussion}

\subsection{The brightness temperature}
From the high-resolution VLBI images, the brightness temperature of
radio core $T_{\rm B}$ in the rest frame can be estimated with
\citep{ghi93}

\begin{equation} T_{\rm B}=\frac{S_{\nu}\lambda^2}{2k\Omega_{\rm
s}}=1.77\times10^{12}(1+z)(\frac{S_{\nu}}{\rm Jy})(\frac{\nu}{\rm
GHz})^{-2}(\frac{\theta_{d}}{\rm mas})^{-2} \end{equation} in which
$z$ is source redshift, $S_{\nu}$ is core flux density at frequency
$\nu$, and $\theta_{\rm d}$ is source angular diameter $\theta_{\rm
d}= a $ with $a$ being the radius of circular components. The
intrinsic brightness temperature $T_{B'}$ can be related with
$T_{B}$ by
\begin{equation} T'_{\rm B}= T_{\rm B}/\delta. \end{equation}
Normally, the upper limit of physically realistic brightness
temperature of nonthermal radio emission can be taken as the
equipartition brightness temperature $T_{\rm in}$ = $5 \times
10^{10}$ K (Readhead 1994).

\begin{figure*}
  \begin{center}
    \includegraphics[height=.3\textheight]{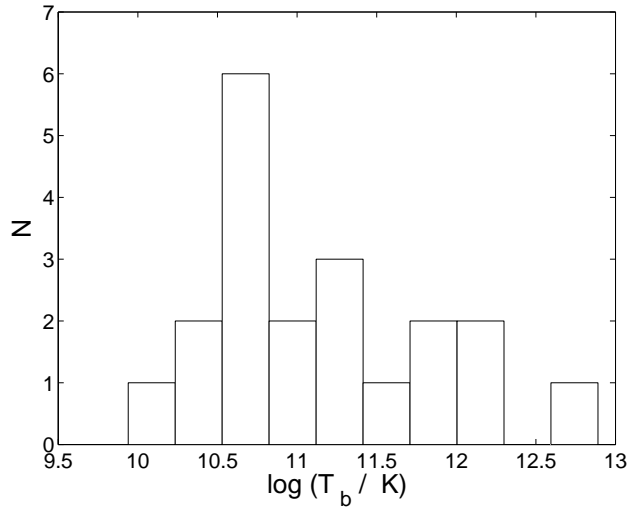}
  \end{center}
\caption{\textbf{The histogram} of brightness temperature of radio cores in
all sources from our observations and all VLBA archive data.
\label{TBdis}}
\end{figure*}

We have calculated the brightness temperature in the source rest
frame for the core components of these nine UHBLs (see Table
\ref{table_model}), in which the measurements from the multi-epoch
and multi-frequency VLBA archive data are also given. The
distribution of all measured brightness temperature is shown in Fig.
\ref{TBdis}. The mean and median values of $T_{\rm b}$ for the radio
core are $\sim10^{11}$ K, which exceeds the equipartition brightness
temperature $T_{\rm in}$. The high brightness temperature suggests
that the beaming effect likely presents in all sources, and it can
be quite strong in some sources, for example, $T_{\rm b}=10^{12.89}$
K in RXS J1458.4+4832, which is even greater than the inverse
Compton catastrophic brightness temperature $10^{12}$ K
\citep{kel69}.

\subsection{The Fermi $\gamma$-ray detection}
The $\gamma$-ray emission of four sources are detected by the Fermi
Large Area Telescope (LAT), and they are listed in the second AGNs
catalog \citep{ackermann2011}, i.e. 2E 0414+0057 with $\gamma$-ray
photon index $\Gamma$=1.98, EXO 0706.1+5913 ($\Gamma$=1.28), RXS
J1012+4229 ($\Gamma$=1.87), RXS J2304+3705 ($\Gamma$=1.96). The
detection of $\gamma$-ray emission provide another evidence of
beaming effect in these sources, in addition to the high brightness
temperature. We note that their $\gamma$-ray photon index $\Gamma$
are all smaller than 2.0. This indicates that the peak frequency of
inverse compton scattering could be higher than the Fermi
$\gamma$-ray frequency, which is consistent with their high
synchrotron peak frequency. Indeed, the photon index are generally
consistent with the mean photon index of HBLs $1.90\pm0.17$
\citep{ackermann2011}, confirming their source classification.

All four sources are resolved with core-jet structure, while only
one source has core-jet structure in the rest five non-Fermi
detected sources. It thus seems that the Fermi detected BL Lac
objects are more likely to have longer jets than non-Fermi sources
in our sample. Although our sample is rather small, it is consistent
with the results shown in \cite{linford11} that the short jet and
point source BL Lacs are less likely to produce $\gamma$-rays. About
75\% of the LAT BL Lacs are classified as long jets, compared to
only 45\% for the non-LAT BL Lacs.


\subsection{The proper motion}
\begin{figure*}
  \begin{center}
    \includegraphics[height=.3\textheight]{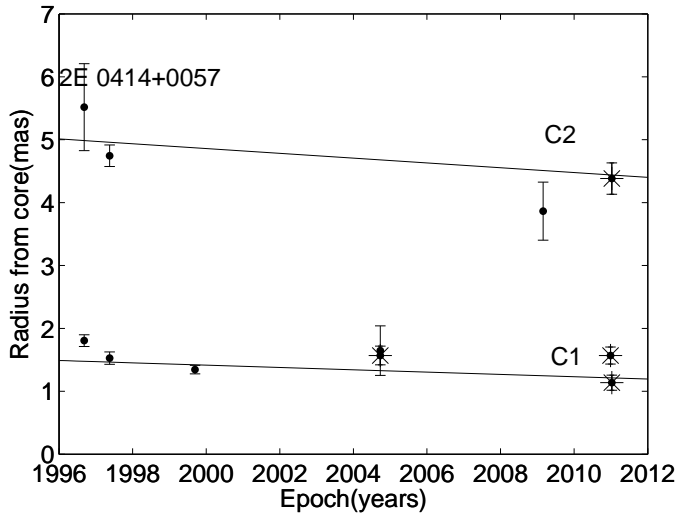}
  \end{center}
\caption{The jet proper motion of 2E 0414+0057. The weighted linear
fits are shown as solid lines for C1 (lower) and C2 (upper). The
solid circles are for data at 5 GHz, and the asterisks at 8 GHz (see
text for details). \label{J0414_comp}}
\end{figure*}

In our nine sources, the proper motion can only be investigated for
two sources 2E 0414+0057 and EXO 0706.1+5913. In 2E 0414+0057, the VLBI observations were collected at eight epochs covering about 14 years (see Table \ref{table_log}). Two jet components are detected and labeled as C1 and C2 (see Fig.
\ref{J0414com}). The relationship of their distance to the radio
core with the observational time is plotted in Fig.
\ref{J0414_comp}. The weighted linear fits of each jet component were
performed in order to calculate the proper motion, which is shown as
solid lines in Fig. \ref{J0414_comp}. The estimated proper motion
are $-0.02\pm0.008 \rm~ mas~ yr^{-1}$ and $-0.038\pm0.021~ \rm mas
~yr^{-1}$ for C1 and C2, respectively. We found that the C1 position
at 8.4 GHz is consistent with that of 5 GHz, with a bit smaller
distance to the core. Therefore, we tentatively combined position
measurements at all frequency to estimate proper motion, however,
this may cause uncertainties in estimating the proper motion, and
can be part reason of negative proper motion. Our results can only
be treated as indicative, not conclusive. But, at least, it shows
that the proper motion is not large in 2E 0414+0057, and jet
components are more like stationary.

In EXO 0706.1+5913, three jet components are identified as C1, C2 and C3 in our EVN image (see Fig.\ref{J0710com} ). To investigate the jet proper motion, the VLBA archive data were collected at four epochs, with the measurements shown in Table \ref{table_model}. At epoch 2010.93, a new component C0' is found at a distance of 1.6 mas to core, which however is not detected in other observations. It may either be resolved due to the higher resolution at 8.6 GHz, or completely a new component. EXO 0706.1+5913 was observed at 1.6 GHz with EVN array at June 7, 2002 \citep{gir06}. The source was resolved to core and one jet component, which is identified as C3, according to its P.A. and distance to the core. From all available data, C1 was measured at three epochs, while only two epochs for C2 and C3 (see Table \ref{table_model}). The proper motion of C1, C2 and C3 are then estimated from multi-epoch data (Fig. 12), with values of $0.113\pm0.115 \rm~ mas~ yr^{-1}$ , $0.638 \rm mas ~yr^{-1}$  and $0.835 \rm mas~yr^{-1}$, corresponding to 0.93c, 5.24c and 6.87c for C1, C2 and C3, respectively. While C1 seems to have subluminal motion, the superluminal motion is only likely detected in C2 and C3, due to the fact that the proper motions of these two components are only estimated from two epochs.

\begin{figure*}
  \begin{center}
    \includegraphics[height=.3\textheight]{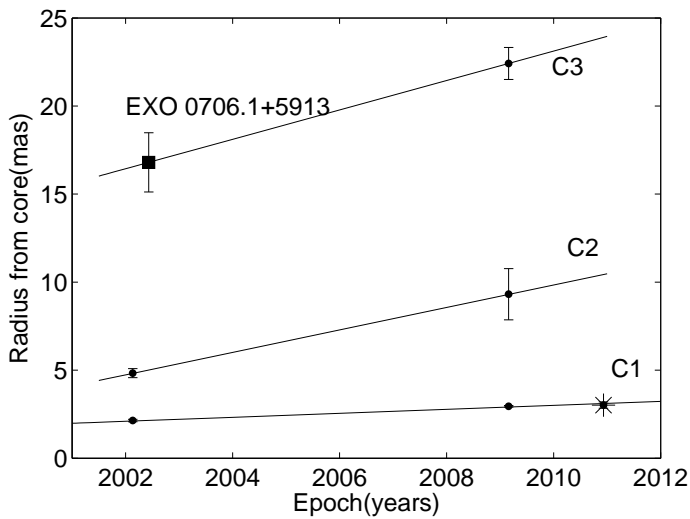}
  \end{center}
\caption{The jet proper motion of EXO 0706.1+5913. The weighted linear
fits are shown as solid lines for C1, C2 and C3. The
solid circles are for data at 5 GHz, and the asterisks at 8 GHz (see
text for details), while the square stands for the 1.6 GHz data component from Giroletti et al.(2006). \label{J0710_comp}}
\end{figure*}
\begin{figure*}
  \begin{center}
    \includegraphics[height=.2\textheight]{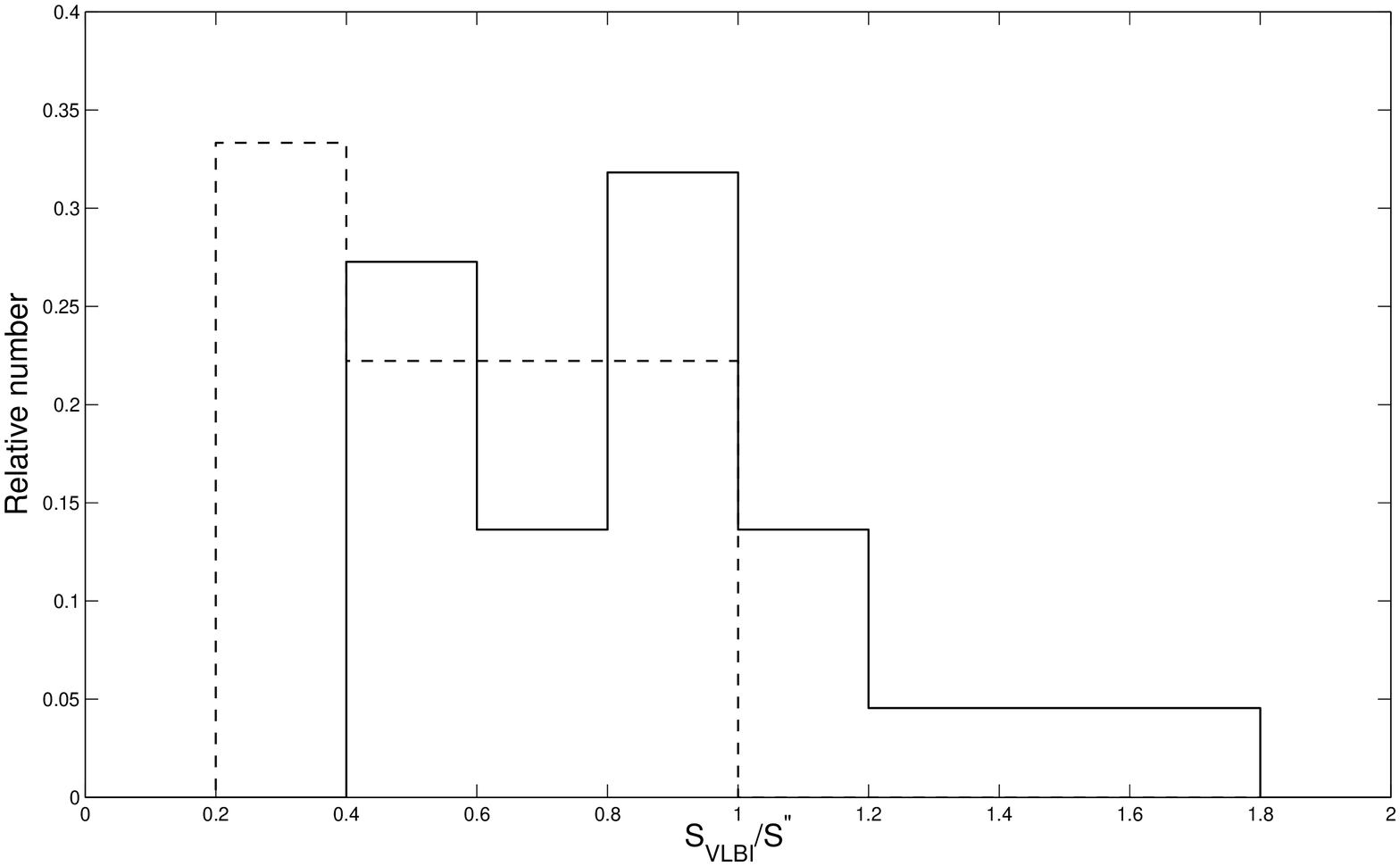}
  \end{center}
\caption{The distribution of $S_{\rm VLBI}$/$S^{''}$ for our UHBLs
and HBLs of Giroletti et al. 2004a. The dashed line indicates the UHBLs, while the real solid lines indicates HBLs}\label{F_ratio}
\end{figure*}
\subsection{The source compactness}
\cite{gir04a} have used a ratio between the arcsecond core flux at 5
GHz $S^{''}$ and the correlated VLBI total flux at the same
frequency $S_{\rm VLBI}$ to show that only a small fraction
($\sim30\%$) of BL Lac objects has a complex subarcsecond structure
invisible in their data. To investigate the compactness and compare
with normal HBLs, we used the MERLIN flux as the lower limit of
arcsecond core flux $S^{''}$, and EVN total flux for $S_{\rm VLBI}$
to calculate the ratio $S_{\rm VLBI}/S^{''}$ for our nine sources.
The $S_{\rm VLBI}/S^{''}$ distributions for our sample is compared
in Fig. \ref{F_ratio} to that of the HBLs sample selected from
\cite{gir04a}. We found that our UHBLs sample has
relatively lower values of $S_{\rm VLBI}/S^{''}$ than HBLs, with
mean and median values of about 0.5 and 0.84, respectively. The
result indicates that our UHBLs sources are less compact than the
normal HBLs, likely due to the less beaming effect.

\subsection{The flux variations}
\begin{figure*}
  \begin{center}
    \includegraphics[height=.3\textheight]{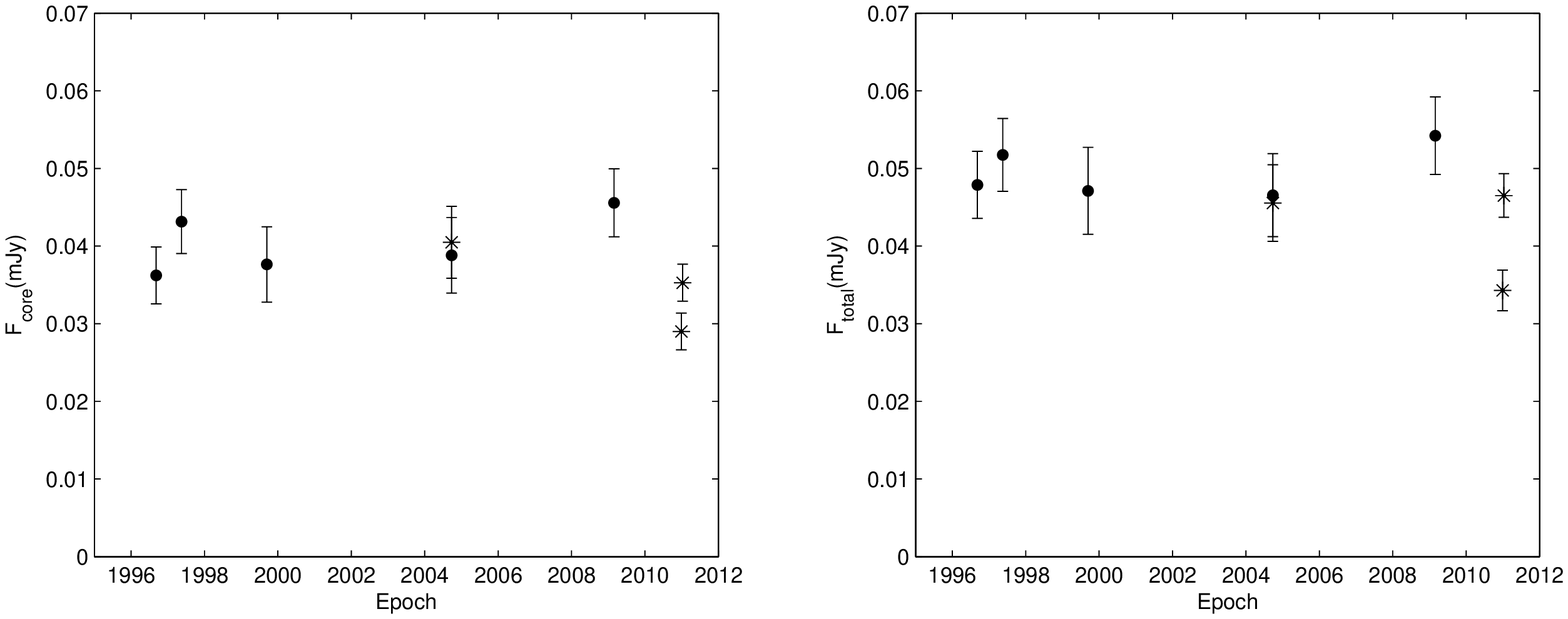}
  \end{center}
\caption{The flux variations in 2E 0414+0057: (left) the core flux,
(right) the total flux. The solid circles and asterisks are for
radio flux at 5 and 8 GHz, respectively. \label{J0414flux} }
\end{figure*}

\begin{figure*}
  \begin{center}
    \includegraphics[height=.3\textheight]{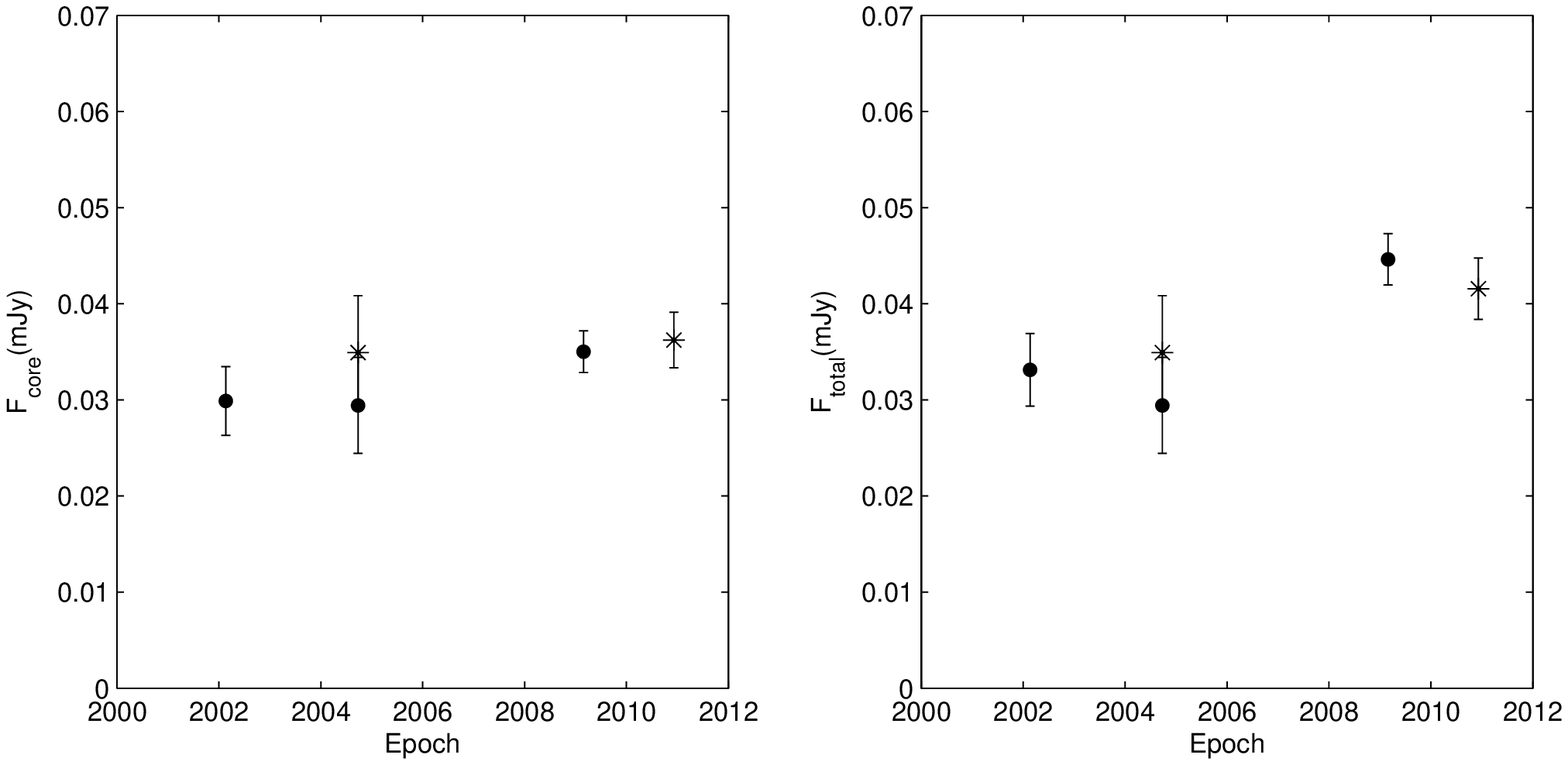}
  \end{center}
\caption{The flux variations in EXO 0706.1+5913: (left) the core flux,
(right) the total flux. The solid circles and asterisks are for
radio flux at 5 and 8 GHz, respectively. \label{J0710_flux} }
\end{figure*}

The multi-epoch VLBI observations of 2E 0414+0057 and EXO
0706.1+5913 are collected to explore the variations both in the core
flux and total flux. The core and total flux density against the
observation time are presented in Figs. \ref{J0414flux} and
\ref{J0710_flux}. The significances of the variations were calculated
using the method presented by DiPompeo et al. (2011),
$$\sigma_{\rm var} = \frac{|S_2-S_1|}{\sqrt{\sigma_2^2+\sigma_1^2}}$$
where $S_1$ and $S_2$ are flux at two epochs, and $\sigma_1$ and
$\sigma_2$ are the corresponding flux uncertainties. We calculated
$\sigma_{\rm var}$ with the fluxes at every two epochs at the same
frequency. The largest $\sigma_{\rm var}$ of 2E 0414+0057 are 2.2
and 3.2 for the core and total flux, respectively, while 1.0 and 2.7
for EXO 0706.1+5913. All the values are smaller than the limit value
of flux variations $\sigma_{\rm var}>4$ given by DiPompeo et al.
(2011). Therefore, there are no significant variations for both the
core and total flux in these two sources, implying that the beaming
effect are likely not severe.
\subsection{UHBLs and HBLs}
Our EVN observations show that the core-jet structure is clearly
detected in five of nine sources. Albeit with a lower detection rate
of jet structure, this is similar to the parsec-scale jet structure
in HBLs and LBLs \citep{gir04a,kharb08,wu06}. No counter-jets are
detected from our observations. All sources are unresolved with
MERLIN, except for J1012+4229, J1319+1405 and J1458+4832, in which
the signs of jet structure can be seen (see Figs. 1 -
\ref{J2304em}).

Normally, it is believed that there are many differences between
LBLs and HBLs, including the multi-band luminosities
\citep{nieppola2006}, the redshift distribution \citep{kharb08}, the
Doppler factors \citep{wu06}, and so on. \cite{kharb08} and
\cite{gir04a} shows that the distribution of total radio power at
1.4 GHz for HBLs and LBLs are significantly different. However, the
differences between LBLs, IBLs, HBLs and UHBLs are less discussed.
To tackle this issue, we collected the data of redshfit, NVSS 1.4
GHz luminosity $L_{\rm NVSS}$ and 5 GHz luminosity $L_{\rm 5~ GHz}$
for the BL Lacs sample in \cite{nieppola2006}, in which each BL Lac
object was classified individually. The differences of UHBLs with
LBLs, IBLs, and HBLs are investigated with Kolmogorov-Smirnov (KS)
test on the distribution of each parameter, and the results are
shown in Table \ref{tablekstest} in which the source numbers of each populations are also shown. We found that the UHBLs and HBLs
have no significant difference in the distribution of $L_{\rm NVSS}$
and redshift, while different in $L_{\rm 5 ~GHz}$. Since 1.4 GHz
radio luminosities are likely less effected by the beaming effect,
therefore, the similar distribution of HBLs and UHBLs indicates that
these two populations may be intrinsically same \cite[see
also][]{wu12}. On the other hand, the different distribution of
$L_{\rm 5 ~GHz}$ implies that the beaming effect could be different,
e.g. a less beaming effect in UHBLs, due to the fact that $L_{\rm 5
~GHz}$ are more influenced by the beaming effect. While the
significant differences are found between UHBLs and LBLs, it's
interesting to note that UHBLs are similar to IBLs according to the
KS test results. Recently, \cite{meyer11} have argued that the IBLs
are more misaligned versions of HBLs with similar jet powers. This
scenario could also be appropriate for UHBLs, at least it is
consistent with the results from direct comparison with HBLs.

Although the high brightness temperature and $\gamma$-ray detection
imply that the beaming effect is likely presented in each UHBLs, the
less compact radio structure, the less variations and proper motion
(although only in several sources) all suggest that the beaming
effect may not be strong, likely less than that of HBLs. \cite{wu06}
found an strong anti-correlation between the Doppler factor and
synchrotron peak frequency.
Since UHBLs are at the high end of peak frequency distribution, they are expected to systematically have lower value of Doppler factor. Our results seem to be consistent with these expectations. A lower Doppler factor may indicate a larger viewing angle in UHBLs as expected from Wu et al. (2007). Alternatively, the Fermi $\gamma$-rays detections in four of nine sources imply that the jet velocity possibly has significantly decreased with respect to the very inner part, however with a small viewing angle generally required by the common large Doppler factors in $\gamma$-ray emission.

Our UHBLs are directly selected form \cite{nieppola2006}. However,
the authors have argued that the actual position of the peak is
probably exaggerated by the use of a parabolic fitting function, the
peak frequencies of these objects cannot be considered as definite.
One extreme possibility is that these sources are completely not
real UHBLs. As shown in \cite{abdo2010}, no evidence was found for
the hypothetical class of UHBLs characterized by a synchrotron
emission that is so energetic to reach the $\gamma$-ray band.
Although it is not definite, we can still see that $\nu_{\rm peak}$
of these sources are apparently higher than normal HBLs according to
their SED in \cite{nieppola2006}. These points might be kept in the
mind when further analyzing our results. The future simultaneous
multi-band data analysis are necessary to explore their source
nature, for example, by model-fitting with synchrotron and
synchrotron self-compton.

\begin{table*}
 \caption{Results of modelfit parameters.}
\label{table_model}
\begin{tabular}{lllllllll}
\hline\hline
Source &Epoch &Freqency &Comp. & Flux density & r  & P.A.& $a$&  log $T_{\rm b}$\\
       &     &  (GHz)    &&(mJy) &            (mas)  &&(mas)&(K)
\\\hline
           2E 0414+0057& 1996.68   &    4.99  &C0 &   36.23    $\pm$   3.66  &     0.00 $\pm$      0.02  &     0.00    &   0.57  $\pm$     0.04  &    10.25\\
   & 1996.68   &    4.99  & C1&    9.80    $\pm$   2.06  &     1.81 $\pm$      0.09  &    56.70    &   1.81  $\pm$     0.19  &     \\
   & 1996.68   &    4.99  & C2&    1.85    $\pm$   1.01  &     5.52 $\pm$      0.69  &   104.26    &   2.84  $\pm$     1.38  &     \\
   & 1997.38   &    4.99  & C0&   43.15    $\pm$   4.11  &     0.00 $\pm$      0.01  &     0.00    &   0.40  $\pm$     0.03  &    10.79\\
   & 1997.38   &    4.99  & C1&    5.81    $\pm$   1.79  &     1.53 $\pm$      0.10  &    71.37    &   1.65  $\pm$     0.20  &     \\
   & 1997.38   &    4.99  &C2 &    2.78    $\pm$   1.39  &     4.75 $\pm$      0.17  &   124.09    &   2.30  $\pm$     0.34  &     \\
   & 1999.70   &    4.99  &C0 &   37.64    $\pm$   4.86  &     0.00 $\pm$      0.02  &     0.00    &   0.54  $\pm$     0.05  &    10.33\\
   & 1999.70   &    4.99  &C1 &    9.48    $\pm$   2.78  &     1.35 $\pm$      0.07  &    80.15    &   1.07  $\pm$     0.13  &     \\
   & 2004.73   &    4.99  &C0 &   38.82    $\pm$   4.87  &     0.00 $\pm$      0.02  &     0.00    &   0.39  $\pm$     0.03  &    10.79\\
   & 2004.73   &    4.99  &C1 &    7.73    $\pm$   2.21  &     1.65 $\pm$      0.39  &    73.29    &   3.90  $\pm$     0.79  &     \\
   & 2004.73   &    8.42  &C0 &   40.49    $\pm$   4.63  &     0.00 $\pm$      0.01  &     0.00    &   0.29  $\pm$     0.02  &    10.75\\
   & 2004.73   &    8.42  &C1 &    5.04    $\pm$   1.72  &     1.57 $\pm$      0.15  &    86.41    &   1.08  $\pm$     0.30  &    \\
   & 2009.16   &    4.99  &C0 &   45.57    $\pm$   4.39  &     0.00 $\pm$      0.01  &     0.00    &   0.33  $\pm$     0.02  &    11.08\\
   & 2009.16   &    4.99  &C2 &    8.64    $\pm$   2.40  &     3.86 $\pm$      0.46  &    90.68    &   3.56  $\pm$     0.92  &    \\
   & 2010.99   &    8.65  &C0 &   29.01    $\pm$   2.36  &     0.00 $\pm$      0.00  &     0.00    &   0.12  $\pm$     0.01  &    11.73\\
   & 2010.99   &    8.65  &C1 &    5.28   $\pm$  1.16  &     1.57 $\pm$      0.14  &    73.87    &  1.38  $\pm$     0.27  &     \\
   & 2011.02   &    8.65  &C0 &   35.27    $\pm$   2.39  &     0.00 $\pm$      0.01  &     0.00    &   0.27  $\pm$     0.01  &    10.72\\
   & 2011.02   &    8.65  &C1 &    9.97    $\pm$   1.39  &     1.14 $\pm$      0.12  &    81.33    &   1.98  $\pm$     0.24  &     \\
   & 2011.02   &    8.65  &C2 &    1.27    $\pm$   0.51  &     4.38 $\pm$      0.25  &    88.15    &   1.46  $\pm$     0.50  &     \\
 EXO 0706.1+5913  & 2002.13   &    5.00  &C0 &   29.89    $\pm$   3.56  &     0.00 $\pm$      0.02  &     0.00    &   0.40  $\pm$     0.04  &    10.56\\
   & 2002.13   &    5.00  &C1 &    1.58    $\pm$   0.87  &     2.14 $\pm$      0.07  &  -128.91    &   0.40  $\pm$     0.14  &     \\
   & 2002.13   &    5.00  &C2 &    1.66    $\pm$   0.93  &     4.83 $\pm$      0.25  &  -156.42    &   1.11  $\pm$     0.51  &     \\
   & 2004.73   &    4.99  &C0 &   29.43    $\pm$   5.00  &     0.00 $\pm$      0.01  &     0.00    &   0.22  $\pm$     0.03  &    11.32\\
   & 2004.73   &    8.42  &C0 &   34.93    $\pm$   5.90  &     0.00 $\pm$      0.01  &     0.00    &   0.13  $\pm$     0.02  &    11.70\\
   & 2009.16   &    4.99  & C0&   35.01    $\pm$   2.16  &     0.00 $\pm$      0.01  &     0.00    &   0.24  $\pm$     0.01  &    11.32\\
   & 2009.16   &    4.99  &C1 &    6.55    $\pm$   0.94  &     2.95 $\pm$      0.05  &  -147.26    &   0.87  $\pm$     0.09  &     \\
   & 2009.16   &    4.99  &C2 &    2.34    $\pm$   1.16  &     9.31 $\pm$      1.45  &  -156.27    &   5.92  $\pm$     2.91  &     \\
   & 2009.16   &    4.99  &C3 &    0.73    $\pm$   0.49  &    22.42 $\pm$      0.91  &  -145.37    &   2.83  $\pm$     1.82  &    \\
   & 2010.93   &    8.65  &C0 &   36.22    $\pm$   2.89  &     0.00 $\pm$      0.00  &     0.00    &   0.16  $\pm$     0.01  &    11.41\\
   & 2010.93   &    8.65  &C0' &    1.80    $\pm$   0.83  &     1.60 $\pm$      0.01  &  -170.87    &   0.17  $\pm$     0.03  &     \\
   & 2010.93   &    8.65  &C1 &    3.54    $\pm$   1.07  &     3.02 $\pm$      0.18  &  -147.79    &   1.31  $\pm$     0.36  &     \\
 1ES 0927+500 & 2009.16   &    4.99  & C0&   12.59    $\pm$   1.24  &     0.00 $\pm$      0.01  &     0.00    &   0.28  $\pm$     0.02  &    10.69\\
   RXS J1012.7+4229& 2009.16   &    4.99  & C0&   19.00    $\pm$   1.42  &     0.00 $\pm$      0.00  &     0.00    &   0.12  $\pm$     0.01  &    11.99\\
   & 2009.16   &    4.99  &C1 &    3.14    $\pm$   0.67  &     3.47 $\pm$      0.06  &     6.41    &   1.35  $\pm$     0.12  &     \\
   & 2009.16   &    4.99  &C2 &    2.62    $\pm$   0.78  &    11.68 $\pm$      0.81  &    25.21    &   5.63  $\pm$     1.62  &     \\
  RGB 1319+140 & 2009.16   &    4.99  & C0&   28.04    $\pm$   2.30  &     0.00 $\pm$      0.01  &     0.00    &   0.35  $\pm$     0.02  &    10.85\\
   & 2009.16   &    4.99  &C1 &    4.44    $\pm$   1.12  &     5.83 $\pm$      0.41  &    77.38    &   3.46  $\pm$     0.81  &    \\
   & 2009.16   &    4.99  &C2 &    0.81    $\pm$   0.41  &    12.43 $\pm$      0.19  &    80.49    &   1.07  $\pm$     0.38  &     \\
  RXS J1341+3959 & 2009.16   &    4.99  &C0 &    6.71    $\pm$   1.06  &     0.00 $\pm$      0.00  &     0.00    &   0.08  $\pm$     0.01  &    12.03\\
  RXS J1410+6100 & 2009.16   &    4.99  &C0 &    7.16    $\pm$   0.90  &     0.00 $\pm$      0.00  &     0.00    &   0.08  $\pm$     0.01  &    12.06\\
 RXS J1458.4+4832   & 2009.16   &    4.99  &C0 &    7.64    $\pm$   0.93  &     0.00 $\pm$      0.00  &     0.00    &   0.05  $\pm$     0.00  &    12.89\\
RXS J2304.6+3705    & 2009.16   &    4.99  &C0 &   10.35    $\pm$   0.95  &     0.00 $\pm$      0.02  &     0.00    &   0.51  $\pm$     0.03  &     9.94\\
   & 2009.16   &    4.99  & C1&    0.90    $\pm$   0.29  &     3.01 $\pm$      0.13  &   -59.25    &   1.15  $\pm$     0.25  &    \\
\hline\hline
\end{tabular}
\vskip 0.1 true cm \noindent
\end{table*}
\begin{table*}

\caption{{\large The VLBI positions of
UHBLs.}\label{table_position}} \center
\begin{tabular}{lccccccccccc}
\hline\hline Source & calibrator &R. A. (h m s) & Dec. (d m s) & R. A. (h m s) & Dec. (d m s) & data \\
                    &          &(J2000)    & (J2000)     &  (J2000)       & (J2000)\\
\hline
1ES 0927+500&J0929+5013&09 30 37.574  & +49 50 25.60  &09 30 37.574 & +49 50 25.549    &EVN\\
RXS 1012.7+4339&J1022+4239&10 12 44.288  &+42 29 57.010 & 10 12 44.305  &+42 29 57.095  &  EVN\\
RGB 1319+140 &J1327+1223& 13 19 31.74 &+14 05 33.140 &13 19 31.742  &+14 05 33.120& EVN\\
RXS J1341.0+3959&J1340+3754&13 41 04.920 &+39 59 35.160 & 13 41 05.108  &+39 59 45.420&MERLIN \\
RXS J1410.5+6100&J1400+6210&14 10 31.700  &+61 00 10.000  &14 10 30.851 &+61 00 12.790&MERLIN \\

RXS J1458.4+4832&J1500+4751&14 58 27.353&+48 32 45.99 & 14 58 27.360  &+48 32 45.979&EVN\\
RXS J2304.6+3705&J2301+3726& 23 04 36.630 & +37 05 07.3  &23 04
36.715 & +37 05 07.422& EVN \\\hline
\end{tabular}
\begin{quote}
\ Column(1): Source name; Column(2): phase referencing calibrator;
Columns (3) - (4): source position (R. A. and Dec.) from NED;
Columns (5) - (6): source position (R. A. and Dec.) measured from
our observations; Column (7) the data used to measure source
position.

\end{quote}
\end{table*}

\begin{table*}

\caption{{\large The KS test on UHBLs with HBLs, IBLs and
LBLs.}\label{tablekstest}} \center
\begin{tabular}{lccccccccccc}
\hline\hline Parameter  & KS statistic & probability & Significantly different & subsets &$N_{subsets}$&$N_{UHBLs}$\\

\hline
$L_{\rm NVSS}$   &0.305  & 0.127  & NO  & HBLs       &  71&  17         \\
                 & 0.265 & 0.253  &  NO & IBLs   & 68&...         \\
                 &0.651  & 7.2328e-06  & YES  & LBLs &   69&...  \\
$z$              & 0.258 &0.228   & NO  & HBLs                &    74&19 \\
                 &0.194  &0.574   & NO  & IBLs                &   73&... \\
                 & 0.3158 &  0.0763  & YES  & LBLs            &    76&... \\
$L_{\rm 5~ GHz}$ &  0.3673 & 0.08   & YES  & HBLs             &     49&14 \\
                 & 0.2653  &0.3708  & NO  & IBLs              &       49&... \\
                 & 0.7255 &  6.3715e-06 & YES  & LBLs         &      51&... \\


\hline
\end{tabular}
\end{table*}


\section{Summary}
We present the EVN and MERLIN observations for nine UHBLs selected
from \cite{nieppola2006}. The VLBA archive data are also combined to
investigate their radio structure and properties. We found that the
core-jet structure is detected in five sources, while four sources
only have compact core on pc scale. The core of all sources show
high brightness temperature (with mean and median values \textbf{log ($T_{\rm b} / {\rm K}) \sim11$}, which implies that the beaming effect likely present in
all sources. When multi-epoch VLBI data are available, we found no
significant variations either for core or total flux density in two
sources (2E 0414+0057 and EXO 0706.1+5913), and no evident proper motion is found in 2E 0414+0057, while the superluminal motion is likely detected in EXO 0706.1+5913. Our sources are found to be less
compact than the typical HBLs in \cite{gir04a}, by
comparing the ratio of the VLBI total flux to the core flux at
arcsec scale. Combining all our results, we propose that the beaming
effect might be present in the jets of UHBLs, however, it is likely
weaker than that of typical HBLs. Moreover, UHBLs could be the less Doppler beamed versions of HBLs with similar jet power. The results are
in good consistence with the expectations from our previous work.

\section*{Acknowledgments}
We thank the anonymous referee for insightful comments
and constructive suggestions. This work is supported by the NSFC grants (No. 10978009, 11163002,
10833002, 11073039, and 10803015), and by the 973 Program (No.
2009CB824800). The European VLBI Network is a joint facility of
European, Chinese, and other radio astronomy institutes funded by
their national research councils. The National Radio Astronomy
Observatory is operated by Associated Universities, Inc., under
cooperative agreement with the National Science Foundation. MERLIN
is a National Facility operated by the University of Manchester at
Jodrell Bank Observatory on behalf of PPARC. This research has made
use of NASA/IPAC Extragalactic Database (NED), which is operated by
the Jet Propulsion Laboratory, California Institute of Technology,
under contract with National Aeronautics and Space Administration.



\begin{thebibliography}{}
\bibitem[Abdo et al.(2010)]{abdo2010}
Abdo, A. A., Ackermann, M.,  Agudo, I., et al. 2010, ApJ, 716, 30

\bibitem[Ackermann et al.(2011)]{ackermann2011}
Ackermann, M., Ajello, M., Allafort, A., et al. 2011, ApJ, 743, 171

\bibitem[Becker, White \& Helfand(1995)]{bec95}
Becker R. H., White R. L., Helfand D. J., 1995, ApJ, 450, 559

\bibitem[Chen et al.(1999)]{chen99} Chen, Y. J., Zhang, F. J., Sjouwerman, L. O. 1999, Ap\&SS, 266, 495

\bibitem[Fossati et al.(1998)]{fossati98}
Fossati, G., Maraschi, L., Celotti, A., Comastri, A., \& Ghisellini,
G. 1998, MNRAS, 299, 433

\bibitem[Gabuzda et al.(2000)]{Gabuzda00}
Gabuzda, D. C., Pushkarev, A. B., Cawthorne, T. V., 2000, MNRAS,
319, 1109

\bibitem[Ghisellini et al.(1999)]{Ghise99}
Ghisellini, G. 1999, ApL\&C, 39, 17

\bibitem[Giommi et al. (2001)]{Giommi01}
Giommi, P., Ghisellini, G., Padovani, P., \& Tagliaferri, G. 2001,
AIPC, 599, 441


\bibitem[Giroletti et al.(2004a)]{gir04a}
Giroletti, M., Giovannini, G., Taylor, G.~B., \& Falomo, R. \ 2004a,
ApJ, 613, 752

\bibitem[Giroletti et al.(2004b)]{gir04b}
Giroletti, M., Giovannini, G., Feretti, L., et al. 2004b, ApJ, 600,
127

\bibitem[Giroletti et al.(2006)]{gir06}
Giroletti, M., Giovannini, G., Taylor, G. B., Falomo, R. 2006, ApJ,
646, 801

\bibitem[Ghisellini et al.(1993)]{ghi93}
Ghisellini G., Padovani P., Celotti A., Marasch L., 1993, ApJ, 407, 65

\bibitem[Ghisellini et al.(1998)]{ghisel98}
Ghisellini, G., Celotti, A., Fossati, G., Maraschi, L. \& Comastri,
A. 1998, \mnras, 301, 451

\bibitem[Jorstad et al.(2001)]{jorstad01}
Jorstad, S.~G., Marscher, A.~P., Mattox, J.~R., et al. 2001, ApJ,
556, 738

\bibitem[Kellermann \& Pauliny-Toth(1969)]{kel69}
Kellermann, K. I., \& Pauliny-Toth, I. I. K. 1969, ApJ, 155, L71

\bibitem[Kharb et al.(2008)]{kharb08}
Kharb, P., Gabuzda, D., Shastri, P. et al. 2008, MNRAS, 384, 230

\bibitem[Linford et al.(2011)]{linford11}
Linford, J. D., Taylor, G. B., Romani, R., et al. 2011, ApJ, 726, 16

\bibitem[Meyer et al.(2011)]{meyer11}
Meyer, E. T., Fossati, G., Georganopoulos, M., et al. 2011, ApJ, 740, 98
1107.5105

\bibitem[Nieppola et al.(2008)]{nieppola2008}
Nieppola, E.,  Valtaoja, E., Tornikoski, M.,  Hovatta, T.,  Kotiranta, M. 2008, A\&A, 488, 867

\bibitem[Nieppola et al.(2006)]{nieppola2006}
Nieppola, E., Tornikoski, M., \& Valtaoja, E. 2006, A\&A, 445, 441

\bibitem[Padovani \& Giommi (1995)]{padovani95}
Padovani, P., \& Giommi, P. 1995, ApJ, 446, 547

\bibitem[Padovani (2007)]{padovani07}
Padovani, P., 2007, AP\&SS, 309, 63

\bibitem[Piner \& Edwards (2004)]{pin04}
Piner, B.~G., \& Edwards, P.~G.\ 2004, ApJ, 600, 115

\bibitem[Piner et al. (2008)]{pin08}
Piner, B. Glenn.,  Pant, Niraj.,  Edwards, Philip G., ApJ, 678, 64P

\bibitem[Rani et al. (2011)]{rani2011}
Rani, Bindu; Gupta, Alok C.; Bachev, R.; Strigachev, A.; Semkov, E.; et al. 2011,
arXiv:1107.0597v1

\bibitem[Readhead(1994)]{rea94}
Readhead A.C.S., 1994, ApJ, 426, 51

\bibitem[Rector et al.(2003)]{rec03}
Rector, T.~A., Gabuzda, D.~C., \& Stocke, J.~T.\ 2003, AJ, 125, 1060



\bibitem[Wu et al.(2007)]{wu06}
Wu, Z. Z., Jiang, D., R., Gu, M. F., Liu, Y., 2007, A\&A, 466, 63

\bibitem[Wu et al.(2009)]{wu09}
Wu, Z. Z.,  Gu, M. F., Jiang, D., R., 2009, RAA, 9, 168

\bibitem[Wu et al.(2012)]{wu12}
Wu, Z. Z.,  Gu, M. F., Jiang, D., R., 2012, submit to APRIM
\end{thebibliography}
\end{document}